\documentclass{article}

\usepackage{arxiv}
\usepackage{amsmath} 

\usepackage[utf8]{inputenc} 
\usepackage[T1]{fontenc}    
\usepackage{hyperref}       
\usepackage{url}            
\usepackage{booktabs}       
\usepackage{amsfonts}       
\usepackage{nicefrac}       
\usepackage{microtype}      
\usepackage{lipsum}		
\usepackage{graphicx}
\usepackage{natbib}
\usepackage{doi}
\usepackage{tikz}
\usepackage[table]{xcolor} 
\usepackage{colortbl}
\usetikzlibrary{shapes.geometric, arrows.meta, positioning} 

\usepackage[table]{xcolor} 
\definecolor{lightred}{RGB}{255,200,200}
\definecolor{lightyellow}{RGB}{255,255,200}
\definecolor{lightgreen}{RGB}{200,255,200}

\title{GRIMM: Genetic stRatification for Inference in Molecular Modeling}

\author{ \href{https://orcid.org/0000-0002-0991-7726}{\includegraphics[scale=0.06]{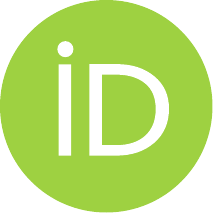}\hspace{1mm}Ashley Babjac} \\
	Department of Marine Sciences\\
	University of Georgia\\
	Athens, GA\\
	\texttt{Ashley.Babjac@uga.edu} \\
	\And
	\href{https://orcid.org/0000-0003-1006-9661}{\includegraphics[scale=0.06]{orcid.pdf}\hspace{1mm}Adrienne Hoarfrost} \\
	Department of Marine Sciences\\
	University of Georgia\\
	Athens, GA\\
	\texttt{Adrienne.Hoarfrost@uga.edu} \\
}

\renewcommand{\shorttitle}{GRIMM: Genetic stRatification for Inference in Molecular Modeling}

\hypersetup{
pdftitle={GRIMM: Genetic stRatification for Inference in Molecular Modeling},
pdfsubject={Enzyme Commission Benchmark Dataset},
pdfauthor={Ashley Babjac and Adrienne Hoarfrost},
pdfkeywords={Enzyme Commission, Sequence Similarity, Swissprot, Benchmarking},
}

\begin{document}
\maketitle

\begin{abstract}
The vast majority of biological sequences on Earth encode unknown functions and bear little resemblance to experimentally characterized proteins, limiting both our understanding of how life interacts with itself and its environment and our ability to harness this functional potential for the bioeconomy. Generalizable prediction of enzyme function from sequence thus remains a central challenge in computational biology. This task is complicated by low sequence diversity of publicly available amino acid sequences relative to environmental proteins for most functional categories and low, imbalanced support across functional labels. 
Models trained and evaluated on these data can inflate model performance metrics and obscure true generalization. To address the limitations of current datasets for training more generalizable function prediction models, we introduce GRIMM (Genetic stRatification for Inference in Molecular Modeling). 
GRIMM is a benchmark for enzyme function prediction that employs genetic stratification, assigning sequence-similarity clusters exclusively to training, validation, or test partitions. Sequences from the same cluster are kept exclusive to train, validation, or test partitions for each label. This genetic stratification approach produces multiple test sets: a closed-set test with the same label distribution as the training set (Test-1), and an open-set test containing novel labels (Test-2), providing a realistic out-of-distribution proxy for discovery of novel enzyme functions. While we apply this genetic stratification approach to construct a benchmark dataset for enzyme function prediction, the method is generally applicable to any sequence-based classification task in which inputs can be clustered by sequence similarity. Importantly, GRIMM formalizes a splitting strategy that is already implicitly used across many biological modeling efforts, providing a unified and reproducible framework for defining closed-set and open-set evaluation based on sequence similarity. The approach is intentionally lightweight, requiring only sequence clustering and label annotations, and can be readily adapted to different similarity thresholds, data scales, and biological tasks without task-specific assumptions. 
This benchmark enables more realistic evaluation of functional prediction models on both familiar and previously unseen functional classes, and establishes a reproducible framework for creating benchmark datasets that more faithfully assess model performance and generalizability on biological prediction tasks.


\end{abstract}


\keywords{Enzyme Commission, Sequence Similarity, Swissprot, Benchmarking}

\section{Introduction}



A persistent challenge in biological sequence modeling is the limited generalizability of Bio-AI models when applied beyond the conditions under which they are evaluated. In practice, sequence-based models predicting biological characteristics from DNA or amino acid sequences are often trained and assessed on datasets that can contain substantial overlap in sequence similarity between training, validation, and test partitions, even though real-world biological applications routinely introduce novel sequences that are evolutionarily distant or absent from the training data. This disconnect between evaluation protocols and deployment conditions leads to inflated performance estimates and precludes accurate assessment of model behavior on truly novel, out-of-distribution (OOD) sequences.
Here in the context of genomics, we specifically use the term out-of-distribution (OOD) to describe protein sequences that differ substantially from the training data in terms of amino acid sequence similarity, and that lie outside the regions of sequence space well represented during model training \cite{shih2025comparative}.

A key contributor to inflated performance estimates is redundancy within commonly used biological sequence datasets. 
Traditional splitting strategies frequently permit homologous sequences to appear across partitions, introducing leakage that reduces the effective difficulty of the prediction task \cite{FerrerFlorensa2024_SpanSeq, shih2025comparative}. As a result, standard benchmarks often fail to capture the challenges associated with generalization to OOD sequences, particularly those occupying sparsely sampled or underexplored regions of sequence space \cite{koh2021wilds}. 



To address these limitations, we introduce GRIMM (Genetic stRatification for Inference in Molecular Modeling), a methodology for constructing similarity-aware train–test splits that yield pseudo–OOD evaluation sets approximating the biological novelty encountered in real-world deployment. In this approach, sequences are grouped by sequence similarity clusters---such as UniRef50 \cite{Suzek2007_UniRef, uniprot2021uniprot} or any user-provided cluster ID---and each cluster is assigned exclusively to one partition for each of the classification labels of interest. This creates a clear separation between training and evaluation sequences that maximizes sequence dissimilarity between sets. We define two test sets: Test-1, a closed-set evaluation containing sequences from the same labels as training; and Test-2, an open-set, pseudo-OOD evaluation containing sequences from labels absent in training derived from orphaned clusters, representing novel regions of sequence space. 
We define Test-2 as ``pseudo'' OOD rather than a true OOD set as it is produced by the clustering procedure of publicly available data rather than true, never-before-seen OOD data. 
We demonstrate GRIMM using the Enzyme Commission (EC) classification system as a concrete example, but the framework is broadly applicable to other structured biological labeling systems, including Gene Ontology terms and protein family annotations. By focusing on the construction of reproducible, pseudo-OOD data splits, GRIMM provides a general framework for evaluating model generalization in biological prediction tasks. Notably, this methodology formalizes practices that researchers may already employ implicitly---such as clustering sequences by similarity to reduce leakage---into a consistent framework with explicit closed-set and open-set definitions. Our methodology exposes limitations of traditional splitting strategies, enables the efficient creation of challenging evaluation partitions, and supports systematic benchmarking of computational methods on both familiar and evolutionarily novel sequences. As part of this work, we also release a five-fold EC functional prediction dataset on amino acid sequences of proteins of experimentally characterized function curated from the SwissProt database \cite{UniProt}, publicly available via GitHub and the HuggingFace API.



\subsection{Related Work}

Several prior studies have emphasized the importance of careful dataset design for evaluating Bio-AI models and protein function prediction specifically. Work on annotation quality and benchmarking has shown that homologous leakage between training and test sets can substantially inflate accuracy and obscure true generalization performance, particularly when random or non-stringent dataset splits are used \cite{Schnoes,Gerlt2016,Radivojac2013,Salzberg2019, FerrerFlorensa2024_SpanSeq}. Related analyses have also highlighted the limitations of homology-based annotation transfer in the “twilight zone” of sequence identity, where alignment signals become unreliable and evolutionary divergence frequently leads to functional shifts \cite{Rost1999,Tawfik2020,Khersonsky2006}. 
In parallel, metagenomic and microbiome studies consistently reveal that the majority of enzyme sequence space remains sparsely characterized, with many protein families occupying remote regions far from well-characterized sequences \cite{Steinegger2019,Zhou2022,Payne2015,Devoto2022}.

Advances in deep learning and protein language models — including UniRep, ProtTrans, and ESM — demonstrate strong in-distribution performance while underscoring the need for standardized evaluation protocols that encourage generalization beyond close homology \cite{UniRep,ProtT5,ESM2,CLEAN}. Many protein language models address sequence redundancy during pretraining through implicit stratification; for example, ESM is trained on UniRef50 clusters to reduce sequence similarity bias. 

More recent efforts have begun to formalize rigorous evaluation tasks that explicitly probe model generalization. For example, the NeurIPS 2024 CARE benchmark suite defines enzyme classification and retrieval tasks using challenging train–test splits designed to evaluate out-of-distribution (OOD) performance relevant to real biological applications \cite{Yang2024CARE}. Similarly, SpanSeq provides a framework for clustering and splitting sequences by sequence similarity for structured or regression-based prediction tasks \cite{FerrerFlorensa2024_SpanSeq}. However, these approaches typically focus on similarity-based splitting without explicitly defining closed-set versus open-set label regimes.



GRIMM is complementary to these efforts: rather than proposing task-specific benchmarks, it provides a general-purpose mechanism for defining closed-set and open-set evaluation that subsumes many of these approaches. In this sense, GRIMM serves as a unifying abstraction over similarity-based splitting strategies that are already widely used but rarely formalized. By enforcing cluster-level separation between training, validation, and test partitions, GRIMM reduces homologous leakage while generating evaluation sets that better reflect the challenges posed by evolutionarily novel or sparsely sampled regions of sequence space. While we use EC functional annotation as a concrete example, the approach is broadly applicable to any labeled sequence dataset for which genetic/protein clustering can be performed, enabling rapid, customizable, and method-agnostic generation of benchmarks for robust training and evaluation of model generalization.

\section{Methodology}


The core principle is \textbf{genetic stratification}: GRIMM accepts sequences clustered by sequence similarity with associated cluster IDs (e.g. UniRef50 IDs), and each cluster within each EC label is assigned exclusively to one partition. This produces two evaluation sets partitioned into two groups:

\begin{itemize}
    \item \textbf{Closed-set sequences (Test-1):} sequences with overlapping labels to training sequences.
    \item \textbf{Open-set sequences (Test-2):} sequences from orphaned clusters with labels absent from training, serving as proxies for novel or sparsely sampled categories (functions).
\end{itemize}


While we illustrate GRIMM using the hierarchical Enzyme Commission (EC) classification system, the methodology is broadly applicable to any biological classification task with associated labels, such as Gene Ontology or protein family assignments whose underlying sequences can be clustered into sequence-similar groups. 
Users may vary sequence identity thresholds, adjust train/validation/test proportions, bin/define custom labels for stratification or apply alternative clustering schemes without modifying the underlying framework. No stratification assumptions beyond cluster exclusivity and underlying labels are required, making GRIMM applicable across a wide range of data regimes and task complexities. The GRIMM benchmark datasets for functional prediction and code are publicly available on \href{https://github.com/Hoarfrost-Lab/GRIMM}{GitHub - code} and via \href{https://huggingface.co/datasets/HoarfrostLab/GRIMM}{Hugging Face - data}. We include both nucleotide and amino acid versions of all splits as well as the source code to recreate the full pipeline and analysis.

\subsection{Source Data}

We used amino acid sequence data from the Universal Protein Resource (UniProt) \cite{UniProt} and associated gene DNA coding sequences in the European Nucleotide Archive (ENA) \cite{europeannucleotidearchive}. The UniProt database is divided into two sections: (i) UniProt/TrEMBL (which includes more sequence diversity but more annotation errors owing to homology-based annotations and error propagation \cite{Schnoes}) and (ii) UniProt/SwissProt (which is carefully curated and provides high confidence accurate functional annotations). For this study, we limited the data retrieved to prokaryotic organisms in SwissProt (May 2025) \cite{UniProt}. 

We mapped UniProt/SwissProt accession numbers to corresponding identifiers in UniRef (Universal Protein Resource Reference Clusters) UniRef50, UniRef90, and UniRef100, which define protein clusters based on amino acid sequence identity at 50, 90, and 100 percent amino acid identity respectively; and to their corresponding EMBL CDS IDs for gene coding sequences in the ENA database \cite{europeannucleotidearchive} using ID mapping files \cite{uniprot2021uniprot, wang2021crowdsourcing} from UniProtKB. Each EMBL CDS ID's corresponding DNA sequence was then obtained from ENA. EC numbers that were either incomplete or missing were removed from the dataset. Individual entries were created for UniProt records with more than one EMBL CDS ID in the nucleotide version of the dataset.

\begin{figure}[t]
\centering
\begin{tikzpicture}[
  node distance=0.7cm and 1.2cm,
  every node/.style={rounded corners, align=center, font=\scriptsize, draw, fill=gray!5, text width=3.0cm, minimum width=2.4cm, minimum height=0.7cm},
  decision/.style={diamond, aspect=2, draw, align=center, fill=blue!5, text width=2.3cm, inner sep=1pt, font=\scriptsize},
  arrow/.style={-Stealth, thick}
]

\node (start) {Retrieve raw sequences \\ + metadata \& cluster mapping};
\node (clean) [below=of start] {Preprocess dataset \\ (filter incomplete labels, normalize sequences)};
\node (group) [below=of clean] {Group by class label \\ \& sequence cluster};
\node (enough) [below=of group, decision] {Sufficient cluster count \\ ($>N$)?};

\node (split) [below left=0.9cm and 2.4cm of enough] {Stratified split by cluster \\ (Train / Val / Test-A) \\ Repeat K×};
\node (few) [below right=0.9cm and 2.4cm of enough] {Insufficient clusters \\ → Low-support groups};

\node (extras) [below=0.5cm of few] {Low-support handling: \\ $\geq$N→Train/Val/Test \ 2→Train/Test \ 1→Singleton};
\node (orphans) [below=0.5cm of extras] {Singleton processing: \\ K-fold split \\ (Train / Test-B)};

\node (end) [below=1.2cm of enough, yshift=-3.5cm] {Final outputs per fold: \\ Train / Validation / Test-A / Test-B};

\draw[arrow] (start) -- (clean);
\draw[arrow] (clean) -- (group);
\draw[arrow] (group) -- (enough);
\draw[arrow] (enough) -| node[near start, above left]{Yes} (split);
\draw[arrow] (enough) -| node[near start, above right]{No} (few);
\draw[arrow] (few) -- (extras);
\draw[arrow] (extras) -- (orphans);

\draw[arrow] (split.south) |- ++(0,-0.7) -| (end.north);
\draw[arrow] (orphans.south) |- ++(0,-0.7) -| (end.east);

\end{tikzpicture}

\caption{Decision-based flowchart for constructing stratified sequence datasets under joint classification and clustering constraints. In our use case, protein sequences are retrieved from SwissProt and grouped by enzyme class (EC number) and sequence-similarity clusters (UniRef50) to prevent cluster overlap across data partitions. Classes with sufficient cluster support are split into training, validation, and in-distribution test sets, while low-support classes are handled separately as ``extras'' or singletons (``orphans''). Singleton classes are used to construct an out-of-distribution evaluation set (Test-2), enabling controlled assessment of generalization beyond well-represented EC classes.}
\label{fig:swissprot_condensed}
\end{figure}
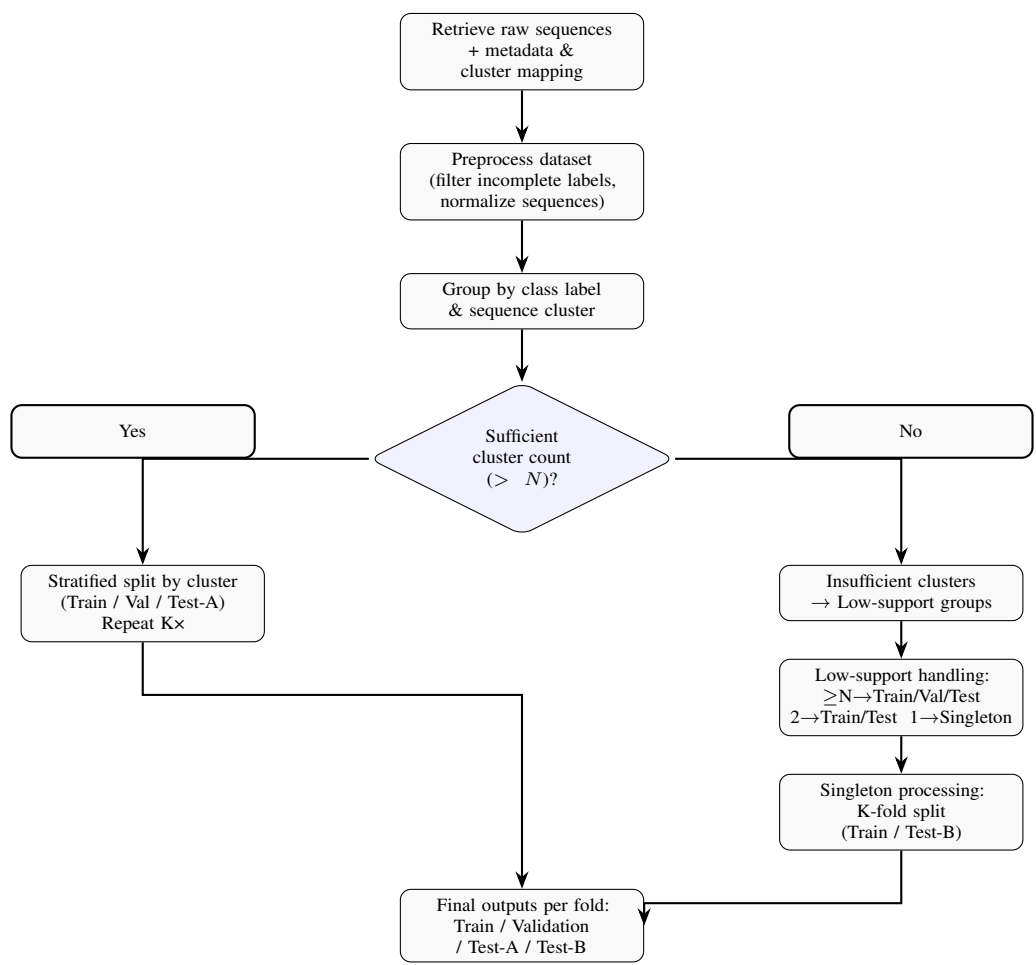

\subsection{Cluster-Based Stratified Splitting}  \label{sec:dataset-methods}

The GRIMM split procedure ensures cluster-level exclusivity across partitions and preserves functional labels (EC numbers) across folds. The basic splitting procedure is as follows: For each unique EC number, we separated sequences into train, valid, and test sets such that no sequence from the same UniRef50 cluster appeared in more than one set. In cases of three or more clusters, we split the cluster IDs roughly 80/10/10 (train/valid/test). In cases where three or fewer UniRef50 clusters were associated with a specific EC number, we added these cluster ids to a separate list of ``extras'' to be processed at the end. We processed these ``extras'' by counting the exact number of available UniRef50 clusters for each EC number. If there were three clusters, we placed the sequences from exactly one cluster in train, one in validation and one in Test-1. If there were two clusters, we placed the sequences from one cluster in train and the other in Test-1. If there was only a singleton cluster, we added this to a separate list of ``orphan'' clusters for downstream processing. We repeat this procedure five times, and shuffle the dataset in between each round in order to have different clusters fall into the train, validation and test sets in each split. 

After finishing the initial 5-fold split, the orphan set is processed. The orphan records represent the proxy for discovery of novel functions, that is, data with unseen ``open-set'' labels that are likely to be underrepresented and more diverse than the training data. We now define  Test-2 which will be comprised of the orphans in each split such that it is open-set and out-of-distribution with respect to both labels and sequences. We include 80\% of orphan clusters in the train (used to preserve data sufficiency but labels do not overlap with Test-2 sequences), and 20\% of orphan clusters in the Test-2 across the 5-fold partition. The final dataset counts of sequences in each split are: train - 185418, validation - 26966, Test-1 - 24617, Test-2 - 420. 

Overall, Test-1 includes EC numbers that are also present in the training data, ensuring consistency for benchmarking---particularly relative to the validation set. By contrast, Test-2 contains EC numbers absent from training, designed to evaluate model generalization to novel or rare functions.




\subsection{Pairwise Sequence Similarity Analysis} \label{sec:seq-identity-methods}

To quantify redundancy and provide insight into sequence overlap across splits, we computed pairwise sequence similarity between training sequences and validation, Test-1 (closed-set), Test-2 (open-set), and external datasets. Test-2 is expected to be more OOD in sequence space than Test-1 due to its lack of label overlap with training. 

\paragraph{External Benchmark Comparison.}
Train sequences were also compared to curated external datasets (\texttt{price-149}, \texttt{new-392}, \texttt{halogenase}), which have varying amounts of overlap to training in label space but are truly OOD in sequence space. These included \texttt{new‑392} (392 enzyme sequences, 177 EC classes), \texttt{price‑149} (149 sequences, 56 EC classes) from the CLEAN benchmark \cite{CLEAN}, and a set of 36 halogenase enzymes from UniProtKB used in prior EC prediction benchmarks \cite{CLEAN,shih2025comparative}, referred to as \texttt{halogenase}.

\paragraph{Pairwise Alignments.}
For each dataset split, all training sequences were compared against sequences in validation, Test-1, Test-2, and external benchmarks. Given datasets of sizes $N_1$ and $N_2$, the number of pairwise comparisons is $N_1 \times N_2$. Alignments and percentage similarity were computed using \texttt{MMseqs2} \cite{mmseqs2} with the BLOSUM62 matrix. Pairs that did not align were assigned a value of zero.

\paragraph{Summary Metrics.}
For each train--dataset comparison, we calculated:

\begin{enumerate}
    \item Number of sequences in training and comparison datasets.
    \item Total possible pairs ($N_1 \times N_2$).
    \item Sum of non-zero percentage similarities.
    \item Mean percentage similarity excluding zeros.
    \item Mean percentage similarity including zeros (treating unaligned sequence pairs as zero).
\end{enumerate}

\paragraph{Cross-Split Aggregation.}
Metrics were computed for each of the five stratified dataset splits (Section~\ref{sec:dataset-methods}) and averaged across splits for each train--dataset pair to obtain robust estimates of redundancy and similarity.

\paragraph{Randomized Baseline.}
To assess whether observed sequence similarity between sets exceeded chance, we generated a randomized baseline using scikit-learn’s \texttt{train\_test\_split} to randomly split sequences across partitions, preserving dataset sizes but ignoring genetic stratification. Pairwise similarities were computed identically, and comparisons were reported as Real, Random, and Diff (Real $-$ Random) values. Negative differences indicate higher sequence similarity in randomized splits relative to genetically stratified splits.

\paragraph{Baseline Performance using CLEAN}

As a demonstration of our dataset against traditional benchmark datasets, we perform a simple analysis using the CLEAN architecture for EC prediction \cite{CLEAN}. We train CLEAN models using default parameters for each of our 5-splits (independently) as well as using \texttt{split100}, then use these models to infer against the true-OOD datasets (\texttt{price-149} and \texttt{new-392}). As a sanity check, we confirm our results from \texttt{split100} and OOD benchmarks match the original CLEAN \cite{CLEAN} (see Table~\ref{tab:generalization_comparison_colored}).

\section{Results}

\paragraph{Genetic Stratification Minimizes Leakage Between Partitions}
GRIMM produces a clear gradient of sequence similarity across dataset partitions (Table~\ref{tab:seq_identity_comparison}). Closed-set evaluation sets (validation and Test-1) exhibit substantial overlap with the training data, with an average percentage similarity excluding zeros of approximately 38–39\%. This is nonetheless a lower percentage similarity than a randomized baseline of >42\%. 

Test-2 sequences, by contrast, are constructed from orphan clusters with labels absent from training and serve as an open-set, out-of-distribution (OOD) proxy representing novel labels and evolutionary distant sequences relative to training. These sequences, along with external benchmark datasets \texttt{price-149} and \texttt{new-392}, show markedly lower average similarities (31–33\% excluding zeros), with the external set \texttt{halogenase} slightly higher at 36\%, reflecting their novelty and minimal sequence similarity to the training set. Including zeros, Test-2 remains nearly identical to \texttt{price-149} and \texttt{new-392} (0.01–0.02, versus ~0.06 for validation and Test-1), reinforcing its role as a stringent pseudo-OOD evaluation set.


Comparison to a randomized baseline highlights that observed sequence similarities in the real data are significantly below chance, for all stratified evaluation sets (Valid, Test-1, and Test-2) but particularly for Test-2, which exhibits -9.16\% less similarity from train compared to a random baseline (Table~\ref{tab:seq_identity_comparison}). External datasets (\texttt{price-149}, \texttt{halogenase}, \texttt{new-392}) are minimally affected by randomization, showing similar values for both including and excluding zeros, which is expected given their role as both open-set and true OOD benchmarks not derived from the same pool of SwissProt sequences.


\paragraph{Genetic Stratification Improves Generalization to OOD Data}
We further compare the performance and generalization of the state-of-the-art function prediction model CLEAN  \cite{CLEAN} when trained on the GRIMM training set compared to the less-stratified dataset \texttt{split100} on which it was originally trained. Poorly generalized models typically display a significant drop in performance from the validation set to a true OOD evaluation set; a more generalizable model should display a smaller performance difference between validation and OOD evaluation set. Additionally, if the GRIMM test sets are sufficiently OOD, models trained on GRIMM should achieve performance on GRIMM test sets that are comparable to external true-OOD evaluation sets.

Critically, the generalization performance of the CLEAN model trained on GRIMM is improved, particularly for Test-2, demonstrating a smaller performance drop between validation and the more challenging open-set test data (Test-2) than observed for CLEAN trained on \texttt{split100} (Table~\ref{tab:generalization_comparison_colored}). CLEAN-\texttt{split100} exhibits a substantial drop from cross-validated F1 on \texttt{split100} sequences to external benchmarks \texttt{new-392} and \texttt{price-149} ($\Delta$F1 = -0.45 to -0.50), whereas CLEAN-GRIMM shows a substantially lower reduction from validation (CV) to one of the OOD datasets (-0.312 \texttt{new-392}), although there is a slightly larger drop for CLEAN-GRIMM on \texttt{price-149} relative to CLEAN-\texttt{split100}. Further, despite the smaller training set size of CLEAN-GRIMM, the model's performance on the external OOD dataset \texttt{new-392} is higher than for CLEAN-\texttt{split100} (0.57 vs 0.452). It is notable that the degree of the performance drop for CLEAN-GRIMM on Test-2 is comparable to that of the external OOD datasets, indicating that it is realistically OOD. Test-1, in contrast, displays a much lower performance drop, consistent with its closer sequence similarity to training. 

The GRIMM stratification approach holds out a significant amount of label and sequence diversity, which is necessary to evaluate generalization but also results in a smaller training set and lower sequence diversity within that training set. This leads to lower model performance overall on CLEAN-GRIMM than the CLEAN-\texttt{split100} model trained on all available data (Table~\ref{tab:generalization_comparison_colored}). 

\begin{table}[t!]
    \centering
    \caption{Comparison of average sequence similarity (\%) between the \textbf{stratified dataset splits (Real)} and a \textbf{randomized baseline (Random)}. Color highlights emphasize closed-set, in-distribution (green), open-set, pseudo-OOD (orange), and external, true-OOD (blue) datasets. The Diff. columns are shaded to indicate magnitude: darker red = larger negative difference.}
    \begin{tabular}{llrrr rrr}
        \toprule
        File1 & File2 &
        \multicolumn{3}{c}{Excl. Zeros} &
        \multicolumn{3}{c}{Incl. Zeros} \\
        \cmidrule(r){3-5} \cmidrule(r){6-8}
        & & Real & Random & Diff. & Real & Random & Diff. \\
        \midrule
        \rowcolor{green!15} train & Valid      & 39.085 & 42.510 & \cellcolor{red!15}-3.425 & 0.0642 & 0.0860 & \cellcolor{red!15}-0.0218 \\
        \rowcolor{green!15} train & Test-1      & 38.836 & 42.664 & \cellcolor{red!20}-3.828 & 0.0651 & 0.0867 & \cellcolor{red!20}-0.0216 \\
        \rowcolor{orange!25} train & Test-2      & 31.824 & 40.984 & \cellcolor{red!70}-9.160 & 0.0180 & 0.0956 & \cellcolor{red!70}-0.0776 \\
        \rowcolor{blue!15} train & Price-149   & 30.034 & 30.387 & \cellcolor{red!10}-0.353 & 0.0179 & 0.0205 & \cellcolor{red!10}-0.0026 \\
        \rowcolor{blue!15} train & New-392     & 33.190 & 33.155 & \cellcolor{red!5} 0.035 & 0.0292 & 0.0333 & \cellcolor{red!5}-0.0041 \\
        \rowcolor{blue!15} train & Halogenase & 36.153 & 36.934 & \cellcolor{red!10}-0.781 & 0.0043 & 0.0062 & \cellcolor{red!10}-0.0019 \\
        \bottomrule
    \end{tabular}
    \label{tab:seq_identity_comparison}
\end{table}

\begin{table}[t]
\centering
\caption{Comparison of F1 performance and generalization between CLEAN trained on GRIMM vs \texttt{split100}. GRIMM-CV is the cross validation performance when aggregated across the 5-fold splits. \texttt{Split100} metrics are as reported by CLEAN \cite{CLEAN}. $\Delta$F1 indicates the differences of the corresponding metric from the CV F1 of the same model. Colors indicate $\Delta$F1 magnitude: green = small drop (defined 0.0-0.2), yellow = moderate (defined 0.2-0.4), red = large drop (defined > 0.4).}
\resizebox{\textwidth}{!}{
\begin{tabular}{r | ccccc | ccc}
\toprule
 & \multicolumn{5}{c}{GRIMM + External} & \multicolumn{3}{|c}{Split100 + External} \\
 \midrule
Metric & CV & Test-1 & Test-2 & New-392 & Price-149 & CV & New-392 & Price-149 \\
\midrule
F1 & 0.882 $\pm$ 0.003 & 0.859 $\pm$ 0.003 & 0.619 $\pm$ 0.004 & 0.570 $\pm$ 0.010 & 0.323 $\pm$ 0.040 & 0.95 & 0.499 & 0.452 \\
$\Delta$F1 & - & \cellcolor{lightgreen}-0.023 & \cellcolor{lightyellow}-0.263 & \cellcolor{lightyellow}-0.312 & \cellcolor{lightred}-0.559& - & \cellcolor{lightred}-0.451 & \cellcolor{lightred}-0.498 \\
\bottomrule
\end{tabular}
}
\label{tab:generalization_comparison_colored}
\end{table}



\section{Discussion}

The stratified dataset and analyses presented here highlight the importance of controlling sequence redundancy and demonstrate the value of formalizing similarity-aware evaluation practices—common in biological sequence modeling but often applied inconsistently or without explicit definitions of label or distributional shifts—when assessing computational approaches. Our genetically informed splitting strategy, based on sequence similarity clusters, results in test partitions with substantially different distributions of sequence similarity to the training data. This mirrors prior work demonstrating that random or insufficiently stringent splits can inflate model performance by permitting homologous sequences to appear in both training and test sets \cite{Schnoes,Gerlt2016,Radivojac2013}. By explicitly constructing Test-1 and Test-2 partitions to maximize sequence dissimilarity from the training set within functional labels, our benchmark more faithfully reflects the evolutionary diversity encountered in downstream, real-world applications such as metagenomic annotation, genome mining, and enzyme discovery.

\paragraph{Stratified splitting results in functionally-relevant sequence divergence}

Our pairwise sequence-similarity analysis confirms that stratification effectively reduces homologous leakage between partitions. Validation and Test-1 sequences exhibit reduced similarity to the training set (approximately 38–39\% similarity excluding zeros vs. 42\% in a random baseline), while Test-2 sequences show substantially lower similarity (approximately 31–33\%). These values fall near or within the ``twilight zone'' of protein homology, where alignment-based inference and homology-based annotation transfer become unreliable \cite{Rost1999}. Rare or biochemically specialized enzyme families, such as halogenases, display even lower similarity, consistent with the observation that many functionally unique enzymes occupy remote or sparsely sampled regions of sequence space \cite{Zhou2022,Payne2015}. Comparisons to randomized baselines further demonstrate that random splits artificially inflate sequence similarity, underscoring the need for stratified dataset design \cite{Salzberg2019}.


\paragraph{Biological implications: models must generalize beyond homology}

These findings have important implications for the interpretation and development of protein language models and their downstream tasks, particularly those, such as enzyme classification, where sequence diversity strongly influences performance. Deep learning models often achieve high accuracy on randomly partitioned datasets \cite{DEEPre,CLEAN,UniRep,ESM2}. However, strong in-distribution performance does not guarantee robust generalization to novel or divergent sequences--an essential requirement for biological discovery and large-scale functional annotation \cite{koh2021wilds}. Reference-free prediction, where models infer function directly from sequence rather than relying on homology-based annotation, represents the greatest potential contribution of these methods. Yet, their current failure to generalize to novel sequences limits this potential, and training and evaluating such models with datasets designed for the explicit goal of generalization is one solution that will contribute to overcoming this limitation.

Our results show that performance drops between validation and test sets correspond closely with the degree of sequence dissimilarity (Table~\ref{tab:generalization_comparison_colored}; Table~\ref{tab:seq_identity_comparison}). Test-1 shows much closer performance to validation that Test-2 and external sets, corresponding with its higher sequence similarity to training ((Table~\ref{tab:seq_identity_comparison}). This confirms that genetic stratification is an effective approach for improving generalization -- but also indicates that creating sufficiently stratified partitions from publicly available data is challenging, with even UniRef50 clusters potentially insufficiently different to produce challenging closed-set tests. This has implications for model training and dataset design in future studies which can be guided by expected sequence dissimilarity during deployment.

Importantly, sequence similarity does not perfectly predict functional similarity. Homologs can diverge in substrate specificity or catalytic mechanism despite high global sequence identity, and small sequence changes in active sites can substantially alter function \cite{Khersonsky2006}. Conversely, distant homologs---or even unrelated proteins with low sequence identity---can catalyze the same reaction or share substrate specificity \cite{Gerlt2016,Tawfik2020,Khersonsky2006}. These phenomena are well-documented in enzyme superfamilies and studies of functional misannotation \cite{Schnoes,Payne2015,Rost1999}. Together, these observations indicate that homology-based transfer alone is insufficient for accurate functional annotation: models must capture local sequence determinants of function rather than relying solely on global similarity.

In applied contexts, such as metabolic engineering, enzyme discovery, and microbiome functional profiling, the ability to annotate sequences with minimal similarity to known proteins is critical. Metagenomic studies continue to reveal that most enzyme diversity remains uncharacterized, with many sequences falling outside well-studied protein families \cite{Steinegger2019,Devoto2022}. This diversity drives biological processes and may contain untapped functional potential for biotechnology. GRIMM’s Test-2 set approximates this challenge by including EC numbers absent from training and sequences with high divergence, while Test-1 contains novel sequences performing familiar functions. As such, GRIMM provides a rigorous test of whether functional prediction models can generalize to underexplored regions of sequence space.

\section{Conclusion}

We have presented a new methodology for constructing genetically stratified benchmark datasets for biological sequences in the context of enzyme function prediction, but extensible to any biological classification task. This methodology is lightweight and can be applied to many similar domains (e.g. Gene Ontologies) with no modification required. Beyond the specific EC benchmark released here, this work makes two broader contributions: (i) it formalizes a general framework for defining closed-set and open-set evaluation in biological sequence modeling based on sequence similarity, reflecting how practitioners already reason about generalization; and (ii) it provides a lightweight, customizable methodology that can be readily adapted across tasks, datasets, and similarity regimes with minimal overhead. By employing stratification of clusters based on sequence similarity, our dataset separates sequences into training, validation, closed-set (Test-1), and open-set (Test-2) datasets, reflecting sequence divergence that more faithfully represents real-world conditions, and providing a more robust framework for assessing model performance across both familiar and novel labels. The resulting benchmark dataset thus enables more accurate evaluation of functional prediction models and highlights the limitations of conventional dataset designs that do not account for sequence redundancy. 
Future efforts will focus on expanding the benchmark to incorporate targeted experimental annotations from broader taxonomic diversity, particularly from underrepresented or uncultured taxa, which will enable training of more generalizable functional prediction models as well as more accurate assessment of model generalization.

\section{Acknowledgments}
This research was developed with funding from the Defense Advanced Research Projects Agency (DARPA) under the DUF Advanced Research Concept. The views, opinions and/or findings expressed are those of the authors and should not be interpreted as representing the official views or policies of the Department of Defense or the U.S. Government.

\bibliographystyle{unsrtnat}
\bibliography{references}  

\end{document}